\documentclass[journal]{IEEEtran}
\usepackage{amsmath,amsfonts}
\usepackage{amsthm}

\usepackage{amssymb}
\usepackage{mathtools}  
\usepackage{bm}   
\newcommand{\Log}{\operatorname{Log}}
\newcommand{\tr}{\operatorname{tr}}
\newcommand{\diag}{\operatorname{diag}}
\usepackage{algorithmic}
\usepackage{array}
\usepackage[caption=false,font=normalsize,labelfont=sf,textfont=sf]{subfig}
\usepackage{textcomp}
\usepackage{stfloats}
\usepackage{cite}
\usepackage{url}
\usepackage{verbatim}
\usepackage{graphicx}
\def\BibTeX{{\rm B\kern-.05em{\sc i\kern-.025em b}\kern-.08em
  T\kern-.1667em\lower.7ex\hbox{E}\kern-.125emX}}
\usepackage{balance}
\begin{document}
\title{State-Space Averaging Revisited via Reconstruction Operators}
\author{Yuxin Yang, Hang Zhou, Hourong Song, Branislav Hredzak
\thanks{This paper is intended to be submitted to IEEE Transactions on Power Electronics}}

\markboth{Letter IEEE Transactions on Power Electronics}%
{Discrete Time modeling for analogue controlled DC-DC converters}

\maketitle

\begin{abstract}
This paper presents an operator-theoretic reconstruction of an equivalent continuous-time LTI model from an exact sampled-data (Poincar\'e-map) baseline of a piecewise-linear switching system. The reconstruction is explicitly expressed via matrix logarithms. By expanding the logarithm of a product of matrix exponentials using the Baker--Campbell--Hausdorff (BCH) formula, we show that the classical state-space averaging (SSA) model can be interpreted as the leading-order truncation of this exact reconstruction when the switching period is small and the ripple is small. The same view explains why SSA critically relies on low-frequency and small-ripple assumptions, and why the method becomes fragile for converters with more than two subintervals per cycle. Finally, we provide a complexity-reduced, SSA-flavored implementation strategy for obtaining the required spectral quantities and a real-valued logarithm without explicitly calling eigen-decomposition or complex matrix logarithms, by exploiting $2\times 2$ invariants and a minimal real-lift construction.
\end{abstract}


\section{Introduction}
\IEEEPARstart{C}{la} ssical state-space averaging (SSA) is widely used in power electronics due to its exceptionally low computational complexity\cite{CUK_SSA}\cite{4421}: it avoids matrix exponentials, avoids discrete-time spectral calculations, and yields compact rational transfer functions. However, SSA is derived via heuristic averaging assumptions that are not obviously consistent with the exact sampled-data dynamics when ripple is non-negligible or when the switching waveform contains multiple subintervals and sign-flipping symmetries.

In contrast, the exact sampled-data model---obtained from the Poincar\'e map of a piecewise-linear (PWL) system---is mathematically unambiguous and can be validated directly against switching simulations over a broad frequency range. This motivates the following question:
\emph{Can we systematically reconstruct a continuous-time LTI surrogate from an exact discrete-time baseline, while retaining SSA-level simplicity whenever the SSA assumptions are valid?}

This paper answers the question by introducing a reconstruction operator based on matrix logarithms. The key observation is that, for a two-subinterval switching system, the logarithm of the one-period state transition matrix admits a BCH expansion whose leading term coincides with the SSA averaged system matrix, and whose higher-order terms are nested commutators that scale with higher powers of the switching period. This viewpoint provides a clean post-hoc justification of SSA in its classical regime, and simultaneously clarifies why SSA becomes unreliable for multi-subinterval systems where BCH truncation becomes structurally fragile.
\section{Exact PWL Poincar\'e Map and the Digital Baseline}
\subsection{PWL dynamics and exact one-period map}
The model in \cite{yang2025competentdiscretetimemodeling} proposed the exact competent model of the switching converter.
In order to facilitate analysis, we expand the model in \cite{yang2025competentdiscretetimemodeling} in detail.\\
Consider a PWL system over one switching period $T_s$, partitioned into $m$ subintervals
$T_1,\dots,T_m$ with $\sum_{i=1}^m T_i=T_s$:
\begin{equation}
\dot{\bm{x}}(t)=A_i\bm{x}(t)+B_i\bm{u},\qquad t\in\mathcal{I}_i.
\end{equation}
Define the exact subinterval flow matrices
\begin{equation}
\begin{aligned}
\Phi_i &\triangleq e^{A_iT_i},\\
\Gamma_i &\triangleq \int_{0}^{T_i} e^{A_i\tau}\,d\tau\, B_i
          =A_i^{-1}(\Phi_i-I)B_i\quad (\text{if }A_i \text{ invertible}).
\end{aligned}
\end{equation}
Then the exact one-period Poincar\'e map (digital baseline) is
\begin{equation}
\begin{aligned}
\bm{x}_{k+1}
&=\Phi\,\bm{x}_k+\Gamma\,\bm{u},\\
\Phi
&\triangleq \Phi_m\Phi_{m-1}\cdots\Phi_1,\\
\Gamma
&\triangleq \Gamma_m+\Phi_m\Gamma_{m-1}+\cdots+\Phi_m\cdots\Phi_2\Gamma_1.
\end{aligned}
\end{equation}

\subsection{Timing (duty) perturbation injection: the exact discrete forcing}
Various models have been proposed to model the discrete dynamics of switching converters.
For a two-subinterval case ($m=2$) with $T_1=(1-D)T_s$, $T_2=DT_s$,
a small timing perturbation $\delta T$ at the switching instant induces an exact
discrete-time forcing of the form
\begin{equation}
\bm{x}_{k+1}
=\Phi\,\bm{x}_k+\Gamma\,\bm{u}
+\bm{\gamma}_{x}\,\delta T + \mathcal{O}(\delta T^2),
\end{equation}
where the forcing direction is
\begin{equation}
\label{eq:gamma_x_exact}
\bm{\gamma}_{x}
=
(A_2-A_1)\bm{X}^\star + (B_2-B_1)\bm{u},
\end{equation}
and $\bm{X}^\star$ is the exact periodic steady state satisfying
$\bm{X}^\star=(I-\Phi)^{-1}\Gamma\bm{u}$.
(Equation \eqref{eq:gamma_x_exact} is the discrete-time counterpart of the duty-injection direction and is the object that later becomes the $B$-matrix column in SSA-like linearizations.)

\section{Reconstruction Operator via Matrix Logarithms}
\subsection{Scaled corrected-IRI convention and the direct term}
We adopt a scaled convention consistent with an exact sampled-data baseline:
a scalar factor $T_s$ can be placed on the discrete-side transfer depending on whether the object is interpreted as a discretization rule or a Z-transform table entry.
In what follows, the reconstructed continuous surrogate is written as
\begin{equation}
\label{eq:Grec}
G_{\mathrm{rec}}(s)
=
\frac{1}{T_s}\,C_c\,(sI-A_c)^{-1}B_c
+
D_{\mathrm{extra}},
\end{equation}
where the constant correction term is chosen as
\begin{equation}
\label{eq:Dextra}
D_{\mathrm{extra}}
\triangleq
D_z-\tfrac12 C_cB_c.
\end{equation}
Importantly, the direct-feedthrough gain $D_c$ of a physical continuous-time system, if present, is \emph{not} scaled by $T_s$; the subtraction in \eqref{eq:Dextra} isolates the discretization-induced alias term from any genuine continuous gain.

```latex

\subsection{Reconstruction operator and its derivation}
\label{subsec:recon_op}

\paragraph{Baseline (given).}
Assume we are given a numerically exact discrete-time \emph{baseline} transfer
\begin{equation}
\label{eq:baseline_Gz}
\tilde G_{\mathrm{tot}}(z)
= C_z\,(zI-A_z)^{-1}B_z + D_z,
\end{equation}
obtained by linearizing the Poincar\'e map of a piecewise-linear (PWL) switching system or a corrected sampled data system using corrected z-transform \cite{yang2025revisitingztransformlaplace}.
Let the sampling period be $T_s$.
Throughout, we adopt the convention that the baseline injection already absorbs $T_s$, i.e.
\begin{equation}
\label{eq:Bz_absorb_Ts}
B_z = T_s\,B_c.
\end{equation}
Moreover, we assume the \emph{phase-aligned} output convention
\begin{equation}
\label{eq:phase_aligned_Cz}
C_z = C_c A_z,
\qquad\text{equivalently}\qquad
C_c = C_z A_z^{-1}.
\end{equation}

\paragraph{Intended continuous surrogate (to reconstruct).}
We seek a continuous-time LTI surrogate
\begin{equation}
\label{eq:Gc_def}
G_c(s)=C_c(sI-A_c)^{-1}B_c + D_c,
\end{equation}
whose \emph{baseband} ($k=0$) interpretation matches \eqref{eq:baseline_Gz}
under the corrected-IRI convention described below.

\paragraph{Corrected-IRI constant term (alias/direct term).}
Even when the original continuous system has no direct feedthrough ($D_c=0$),
the corrected impulse-response-invariance (IRI) interpretation produces
an unavoidable constant term in the \emph{discrete} baseline:
\begin{equation}
\label{eq:Dz_alias_term}
D_z^{(\mathrm{alias})}
= T_s\cdot \frac{1}{2}\,C_cB_c.
\end{equation}
Therefore, in general, the baseline constant satisfies
\begin{equation}
\label{eq:Dz_split}
D_z = T_s\cdot \frac{1}{2}\,C_cB_c + D_c,
\qquad\Longrightarrow\qquad
D_c = D_z - T_s\cdot\frac{1}{2}\,C_cB_c.
\end{equation}

\paragraph{Reconstruction operator.}
Choose a matrix-logarithm branch $\Log(\cdot)$ (or an approximation via BCH, cf.\ below),
and define the reconstruction operator $\mathcal{R}$ mapping the baseline
$(A_z,B_z,C_z,D_z)$ into a continuous surrogate $(A_c,B_c,C_c,D_c)$ by
\begin{equation}
\label{eq:recon_operator_final}
\boxed{
\begin{aligned}
A_c &\triangleq \frac{1}{T_s}\,\Log(A_z),\\[0.2em]
B_c &\triangleq \frac{1}{T_s}\,B_z,\\[0.2em]
C_c &\triangleq C_zA_z^{-1},\\[0.2em]
D_c &\triangleq D_z - T_s\cdot \frac{1}{2}\,C_cB_c.
\end{aligned}}
\end{equation}
With \eqref{eq:Bz_absorb_Ts}--\eqref{eq:Dz_split}, the mapping \eqref{eq:recon_operator_final}
is dimensionally consistent and explicitly removes the corrected-IRI alias/direct contribution.

\subsubsection{Derivation of \eqref{eq:recon_operator_final}}
\label{subsubsec:recon_derivation}

\paragraph{Step 1: reconstruct $A_c$ (logarithm or BCH).}
By construction of the baseline Poincar\'e map, $A_z$ is the discrete state-transition
over one sampling interval. We therefore enforce
\begin{equation}
\label{eq:Az_exp}
A_z = e^{A_cT_s}
\qquad\Longrightarrow\qquad
A_c = \frac{1}{T_s}\,\Log(A_z),
\end{equation}
where $\Log(\cdot)$ denotes a chosen branch.

\paragraph{Step 2: $B_z$ already absorbs $T_s$.}
In our baseline convention, the injection is scaled as in \eqref{eq:Bz_absorb_Ts},
hence
\begin{equation}
\label{eq:Bc_from_Bz}
B_c = \frac{1}{T_s}\,B_z.
\end{equation}

\paragraph{Step 3: phase-aligned $C_z$ implies $C_c=C_zA_z^{-1}$.}
Under the phase-aligned convention \eqref{eq:phase_aligned_Cz},
\begin{equation}
\label{eq:Cc_from_Cz}
C_z = C_cA_z
\qquad\Longrightarrow\qquad
C_c=C_zA_z^{-1}.
\end{equation}

\paragraph{Step 4: corrected-IRI constant term.}
The corrected-IRI baseline contains an unavoidable alias/direct term
\eqref{eq:Dz_alias_term} even if $D_c=0$; thus \eqref{eq:Dz_split} holds and yields
\begin{equation}
\label{eq:Dc_from_Dz}
D_c = D_z - T_s\cdot\frac{1}{2}\,C_cB_c.
\end{equation}
Combining \eqref{eq:Az_exp}--\eqref{eq:Dc_from_Dz} gives \eqref{eq:recon_operator_final}.

\subsubsection{BCH-based approximation of the logarithm}
\label{subsubsec:recon_bch}

In many PWL switching models, the baseline transition admits a factorization
\begin{equation}
\label{eq:Az_factorization}
A_z = e^{A_mT_m}\cdots e^{A_2T_2}e^{A_1T_1},
\qquad \sum_{i=1}^m T_i = T_s,
\end{equation}
where $(A_i,T_i)$ are the continuous matrices and durations of the PWL subintervals.
To avoid forming $\Log(A_z)$ directly, one may approximate
\begin{equation}
\label{eq:Omega_def}
A_z = e^{\Omega},
\qquad
A_c \approx \frac{1}{T_s}\,\Omega,
\end{equation}
where $\Omega$ is obtained via a truncated Baker--Campbell--Hausdorff (BCH) expansion.
For example, for two factors $A_z=e^{X}e^{Y}$ with $X=A_1T_1$ and $Y=A_2T_2$,
a second-order BCH truncation gives
\begin{equation}
\label{eq:BCH2}
\Omega \approx X+Y+\frac12[Y,X],
\qquad [Y,X]\triangleq YX-XY,
\end{equation}
hence
\begin{equation}
\label{eq:Ac_BCH2}
A_c^{(\mathrm{BCH2})}
\triangleq \frac{1}{T_s}\left(X+Y+\frac12[Y,X]\right).
\end{equation}
Higher-order truncations and multi-factor generalizations of \eqref{eq:BCH2}
may be used when needed; the reconstruction of $(B_c,C_c,D_c)$
still follows \eqref{eq:recon_operator_final}.

\section{BCH View: Why SSA Works (and When It Fails)}
\subsection{Two-subinterval case: $\Log(e^{A_2T_2}e^{A_1T_1})$}
For $m=2$,
\begin{equation}
\Phi=e^{A_2T_2}e^{A_1T_1}.
\end{equation}
Let $X\triangleq A_1T_1$ and $Y\triangleq A_2T_2$. Then
\begin{equation}
\label{eq:BCH}
\begin{aligned}
\Log(\Phi)
&=\Log(e^Ye^X)\\
&=X+Y+\tfrac12[Y,X]
+\tfrac{1}{12}\bigl([Y,[Y,X]]+[X,[X,Y]]\bigr)+\cdots,
\end{aligned}
\end{equation}
where $[Y,X]\triangleq YX-XY$ is the commutator.
Therefore the reconstructed continuous matrix is
\begin{equation}
\label{eq:Ac_BCH}
\begin{aligned}
A_c
&=\frac{1}{T_s}\Log(\Phi)\\
&=\frac{1}{T_s}(A_1T_1+A_2T_2)
+\frac{1}{2T_s}[A_2T_2,A_1T_1]
+\mathcal{O}(T_s^2).
\end{aligned}
\end{equation}

\subsection{SSA emerges as the leading BCH truncation}
Using $T_1=(1-D)T_s$ and $T_2=DT_s$,
the leading term of \eqref{eq:Ac_BCH} becomes
\begin{equation}
\label{eq:SSA_Aavg}
\begin{aligned}
A_{\mathrm{avg}}
&\triangleq (1-D)A_1 + DA_2
=\frac{1}{T_s}(A_1T_1+A_2T_2),
\end{aligned}
\end{equation}
which is exactly the SSA averaged system matrix.
Similarly, if $\bm{X}^\star$ is close to the averaged equilibrium
$\bm{X}_{\mathrm{avg}}^\star$ (small ripple),
then the exact duty injection direction \eqref{eq:gamma_x_exact}
approximates the SSA duty injection direction
\begin{equation}
\label{eq:SSA_Bd}
\bm{b}_d^{\mathrm{SSA}}
=
(A_2-A_1)\bm{X}_{\mathrm{avg}}^\star + (B_2-B_1)\bm{u}.
\end{equation}
This explains why the $B$-matrix column obtained from a BCH-based reconstruction
often appears ``the same'' as that obtained from SSA: they share the same structural form,
and differ mainly through how accurately the operating point $\bm{X}^\star$ is computed.

\subsection{Why the higher-order BCH terms are small in the SSA regime}
Assume $\|A_i\|\le M$ and $T_i=\mathcal{O}(T_s)$.
Then $\|X\|,\|Y\|=\mathcal{O}(T_s)$ and
\begin{equation}
\|[Y,X]\|\le 2\|Y\|\,\|X\|=\mathcal{O}(T_s^2).
\end{equation}
Consequently,
\begin{equation}
\frac{1}{2T_s}[Y,X]=\mathcal{O}(T_s),
\end{equation}
and the nested commutators in \eqref{eq:BCH} contribute
$\mathcal{O}(T_s^2)$ and smaller to $A_c$.
Thus, when $T_s$ is small (high switching frequency) and the ripple is small so that
$\bm{X}^\star\approx \bm{X}_{\mathrm{avg}}^\star$,
SSA corresponds to a controlled truncation of the exact reconstruction.

\subsection{Why multi-subinterval converters naturally break SSA/BCH simplicity}
When $m\ge 3$,
\begin{equation}
\Phi=e^{A_mT_m}\cdots e^{A_2T_2}e^{A_1T_1},
\end{equation}
and $\Log(\Phi)$ becomes a multi-factor BCH expansion.
Even if each factor is close to identity, the number of cross commutators and nested
commutators grows rapidly and the cancellation patterns become topology-dependent.
As a result, (i) the leading ``averaged'' term is no longer the whole story, and
(ii) truncation may be structurally biased rather than merely quantitatively small.
This provides a principled explanation for the empirical observation that SSA is robust
for classical two-subinterval PWM converters but becomes fragile for multi-edge/multi-subinterval switching patterns.

\section{A Symbolic Two-State Example with Sign Symmetry: SSA-Level Complexity}
\subsection{Problem structure}
Consider a two-state map that includes a diagonal sign matrix
\begin{equation}
\label{eq:Phi_sign}
\Phi = D_r\,e^{\Omega},
\qquad
D_r=\diag(-1,1),
\qquad
\Omega\in\mathbb{R}^{2\times 2}.
\end{equation}
Such a structure is common in rectified/half-cycle constructions and implies
$\det(\Phi)=\det(D_r)e^{\tr(\Omega)}<0$, hence $\Phi$ has one positive and one negative real eigenvalue.

\subsection{No eig, no logm: eigenvalues from invariants}
For any $2\times 2$ matrix, eigenvalues satisfy
\begin{equation}
\lambda_{1,2}(\Phi)
=
\frac{\tr(\Phi)\pm\sqrt{(\tr\Phi)^2-4\det(\Phi)}}{2}.
\end{equation}
Thus, it suffices to obtain $\tr(\Phi)$ and $\det(\Phi)$.
First, $\det(\Phi)$ is immediate:
\begin{equation}
\label{eq:detPhi}
\det(\Phi)=\det(D_r)\det(e^{\Omega})=\det(D_r)\,e^{\tr(\Omega)}=-e^{\tr(\Omega)}.
\end{equation}
Second, compute $\tr(\Phi)=\tr(D_re^{\Omega})$ using the closed form of $e^{\Omega}$
in terms of invariants:
let $\mu\triangleq \tfrac12\tr(\Omega)$ and $\Delta^2\triangleq \mu^2-\det(\Omega)$.
Then
\begin{equation}
\label{eq:exp2x2}
\begin{aligned}
e^{\Omega}
&=
e^{\mu}
\Bigl(
\cosh\Delta\,I
+
\frac{\sinh\Delta}{\Delta}\,(\Omega-\mu I)
\Bigr),
\end{aligned}
\end{equation}
(with the standard $\sin/\cos$ substitution when $\Delta$ is purely imaginary).
Because $\tr(D_r)=0$, we obtain
\begin{equation}
\label{eq:trPhi}
\begin{aligned}
\tr(\Phi)
&=\tr(D_re^{\Omega})\\
&=
e^{\mu}\,\frac{\sinh\Delta}{\Delta}\,\tr(D_r\Omega).
\end{aligned}
\end{equation}
Equations \eqref{eq:detPhi}--\eqref{eq:trPhi} provide $\lambda_{1,2}(\Phi)$ with only:
(i) traces/determinants of $2\times 2$ matrices, and
(ii) scalar $\exp,\sinh,\cosh$ evaluations---an SSA-level complexity profile.

\subsection{Real logarithm via minimal real-lift}
Because $\Phi$ has a negative eigenvalue, a real $2\times 2$ logarithm may not exist
on the desired branch. A minimal remedy is to perform a real-lift:
let $\lambda_{-}<0$ be the negative eigenvalue and define
\begin{equation}
S_{\mathrm{ext}} \triangleq \mathrm{blkdiag}(\Phi,\lambda_{-})\in\mathbb{R}^{3\times 3}.
\end{equation}
Then a real logarithm can be constructed using a $2\times 2$ rotation-log block for $\lambda_{-}$:
\begin{equation}
\label{eq:realLogBlock}
\frac{1}{T_s}
\begin{bmatrix}
\log|\lambda_{-}| & -\pi\\
\pi & \log|\lambda_{-}|
\end{bmatrix},
\end{equation}
together with $\frac{1}{T_s}\log(\lambda_{+})$ for the positive eigenvalue.
This yields a real $A_c$ satisfying $e^{A_cT_s}=S_{\mathrm{ext}}$.
Crucially, the accompanying $B,C,D$ can be kept aligned with the original state choice
by embedding $B$ and $C$ with a zero in the lifted dimension, and by using the
phase-consistent relation $C_c=C_zS_{\mathrm{ext}}^{-1}$.

\subsection{What this buys us}
The above procedure preserves the exact discrete-time baseline structure
while reducing the most expensive steps:
no eigen-decomposition is required to get $\lambda_{\pm}$,
no complex matrix logarithm is needed,
and the only nontrivial transcendental evaluations are scalar.
This matches the spirit of SSA (simple algebraic steps) while retaining a systematic
connection to an exact sampled-data model.

\section{Summary}
SSA can be interpreted as the leading-order truncation of a matrix-logarithm-based
reconstruction of the exact sampled-data baseline. The BCH expansion clarifies both
the success of SSA in the high-frequency/small-ripple regime and the structural reasons
it becomes unreliable for multi-subinterval switching patterns. By exploiting $2\times 2$
invariants and a minimal real-lift, the reconstruction can be implemented with SSA-like
computational complexity while remaining tightly tied to the exact Poincar\'e map.

\bibliographystyle{IEEEtran}
\bibliography{IEEEabrv,biblio}

\end{document}